\documentclass[11 pt]{article}
\def\be{\begin{equation}}
\def\eea{\end{eqnarray}}
\def\bea{\begin{eqnarray}}
\def\ee{\end{equation}}

\usepackage[psamsfonts]{amssymb}
\usepackage{amsmath}
\author{F. Kheirandish$^{1}$ \footnote{fardin$_{-}$kh@phys.ui.ac.ir} and M.
Amooshahi$^{1}$ \footnote{amooshahi@sci.ui.ac.ir}
\\ $^{1}$ {\small Department of Physics, University of Isfahan,}
\\ {\small Hezar Jarib Ave., Isfahan, Iran.}}
\title{Quantum Charged Non-Linear Nano-String and Quantum Vacuum}
\begin{document}
\maketitle
\begin{abstract}
\noindent The classical and quantum  dynamic of a nonlinear
chareged vibrating string and its interaction with quantum vacuum
field is investigated. Some probability amplitudes for transitions
between vacuum field and quantum states of the string are
obtained. The effect of nonlinearity on some probability
amplitudes is investigated and finally the corect equation for
string containing the vacuum and radiation reaction field is
obtained.
\end{abstract}
\section{Introduction}
In QED a charged particle in quantum vacuum interacts with the
vacuum field and its own field known as radiation reaction. In
classical electrodynamics there is only the radiation reaction
field that acts on a charged particle in the vacuum. The vacuum
and radiation reaction fields have a fluctuation-dissipation
connection [1]and both are required for the consistency of QED.
For example the stability of the ground state, atomic transitions
and lamb shift can only be explained by taking into account both
fields. If self reaction was alone the atomic ground state would
not be stable[1,2]. In some cases the self reaction effects can
be derived equivalently from the corresponding classical
radiation theory [3]. When a quantum mechanical system interacts
with the quantum vacuum of electromagnetic field, the coupled
Heisenberg equations for both system and field give us the
radiation reaction field. For example it can be shown that the
radiation reaction for a charged harmonic oscillator is
$\frac{2e^2}{3c^3}$,[1,2]. One method for generating coherent
states is by interacting a classical current source with the
quantized electromagnetic field, where the probability $ P_n$ for
emission of n photons when neither the momenta nor the
polarization are observed is a poissonian distribution [4]. In
this paper,we investigate the quantum dynamics of a non-linear
nano-scale charged string and it's interaction with the quantum
vacuum field. The effect of nonlinear term on probability
amplitude for some transitions are investigated. In the last
section the radiation reaction field for the quantized vibrating
string is derived and it is shown that the correct equation for
string contains the vacuum as well as the radiation
\section{ Quantum nonlinear dynamic of a nano vibrating charged string.}
Consider a nano-string  with lenght $L$  on $x$-axis
  and let's apply the periodic boundary conditions. The mechanical wave function
  $y(x,t) $ of this string satisfies the following non-linear equation
\begin{equation}\label{f36.5}
\frac{\partial^2 y}{\partial t^2}-v^2\frac{\partial^2 y}{\partial
x^2 }=\frac{\gamma}{2}\frac{\partial^2 y}{\partial
x^2}(\frac{\partial y}{\partial x})^2,
\end{equation}
where $ \gamma $ and $ v $ are constant depend on the string .this
equation can be obtained from the following Lagrangian density
\begin{equation}\label{f36}
\pounds=\frac{1}{2}( \frac{\partial y}{\partial
t})^2-\frac{1}{2}v^2( \frac{\partial y}{\partial
x})^2-\frac{\gamma}{24}( \frac{\partial y}{\partial x})^4,
\end{equation}
the canonical momentum density corresponding to $ y(x,t)$ is
\begin{equation}\label{f37}
\pi_y(x,t)=\frac{\partial\pounds}{\partial(\partial_ty)}=\frac{\partial
y(x,t)}{\partial t}
\end{equation}
and the canonical quantization rule is [5]
\begin{equation}\label{f38}
[\pi_y(x,t),y(x',t)]=-i\delta(x-x')
\end{equation}
The lagrangian density (\ref{f36})give us the Hamiltonian
\begin{equation}\label{f39}
H_s(t)=\int_{0}^{L}
dx(\frac{\pi_y^2}{2}+\frac{1}{2}v^2(\frac{\partial y}{\partial
x})^2+\frac{1}{24}\gamma (\frac{\partial y}{\partial x})^4).
\end{equation}
By expanding $y(x,t)$ and $ \pi_y(x,t) $ in terms of orthogonal
periodic functions $ e^{\frac{i2\pi nx}{L}}$
\begin{equation}\label{f41}
y(x,t)=\sum_{n=-\infty}^{+\infty}C_n(t) e^{\frac{i2\pi
nx}{L}}\hspace{1.00
cm}\pi_y(x,t)=\sum_{n=-\infty}^{+\infty}\dot{C_n}(t)
e^{\frac{i2\pi nx}{L}}
\end{equation}
by using of (\ref{f38}) we find the following commutation
relations
\begin{equation}\label{f42}
[C_n(t),\dot{C_m}(t)]=\frac{i}{L}\delta_{n,-m}\hspace{1.00 cm}
[C_n(t),\dot{C_m}^\dag(t)]=\frac{i}{L}\delta_{n,m},
\end{equation}
because of hermiticty of operators $ y(x,t) $ and $ \pi_y(x,t) $
we find from (\ref{f41}),
\begin{equation}\label{f42.5}
C_n^\dag(t)=C_{-n}(t),\hspace{1.00cm}
\dot{C}_n^\dag(t)=\dot{C}_{-n}(t),
\end{equation}
now we rewrite the Hamiltonian (\ref{f39}) in terms of $C_{n}$'s
\begin{eqnarray}\label{f43}
&&H_s(t)=H_{1s}+H_{2s}\nonumber\\
&&H_{1s}=\frac{L}{2}\sum_{n=-\infty}^{+\infty}
\dot{C_n}(t)\dot{C_n}^\dag+\frac{L}{2}\sum_{n=-\infty}^{+\infty}\omega_n^2C_nC_n^\dag\nonumber\\
&&H_{2s}=-\sum_{n,s,t=-\infty}^{+\infty}M_{n,s,t}
C_{n}C_{s}C_{t}C_{n+s+t}^\dag\nonumber\\
&&M_{n,s,t}=\frac{2\pi^4 \gamma}{3L^3}nst(n+s+t)
\end{eqnarray}
where $ \omega_n=\frac{2\pi v n}{L}$ . Hamiltonian (\ref{f43})give
the Heisenberg equations for $ C_j $'s
\begin{eqnarray}\label{f44}
&&\ddot{C}_j+\omega_j^2C_j=\sum_{s,t=-\infty}^{+\infty}L_{s,t}C_{s}C_{t}C_
{s+t-j} ^\dag\nonumber\\
&&L_{s,t}=-\frac{8\pi^4 \gamma j}{3L^4}st(s+t-j)
\end{eqnarray}
Let us define the operator $\widetilde{O}(t)$, for any operator $
O^s $ in schrodinger picture as
\begin{equation}\label{f44.1}
\widetilde{O}(t)=e^{iH_{1s}t}O^s e^{-iH_{1s}t},
\end{equation}
so in the special case for the operators $ C_{n}(t) $ we have
\begin{equation}\label{f44.2}
\widetilde{C}_{n}(t)=e^{iH_{1s}t}C_{n}(0)e^{-iH_{1s}t}=B_{-n}^\dag
e^{i\omega_{n}t}+B_{n}e^{-i\omega_{n}t}.
\end{equation}
where we have used from (\ref{f42.5}). New operators $B_{n}$ and
$B_{-n}^{\dag}$, are annihilation and creation operators of
phonons of type $|n\rangle$, i.e, with the wave function
$\frac{1}{\sqrt{L}}e^{(\frac{2\pi n}{L}x)}$ and satisfy the
following commutation relations
\begin{equation}\label{f44.3}
[B_{n},B_{m}^\dag]=\frac{\delta_{n,m}}{2L\omega_{n}}.
\end{equation}
A state with  $j$  phonons in fock space with the corresponding
modes, $s_1, s_2,...,s_j$, is denoted by $|s_1,s_2,...,s_j>$,
such that
\begin{eqnarray}\label{f44.4}
&&B_{m}|s_1,s_2,...s_j\rangle=\frac{1}{\sqrt{2L\omega_{m}}}\sum_{r=1}^j
\delta_{m,s_r} |s_1,...,s_{r-1},s_{r+1},...s_j\rangle,
\nonumber\\
&&B_{m}^\dag|s_1,s_2,...s_j\rangle=\frac{1}{\sqrt{2L\omega_{m}}}|m,s_1,...s_j\rangle,
\end{eqnarray}
substituting $ C_{n}(0)=\widetilde{C}_{n}(0) $ from (\ref{f44.2})
in (\ref{f43}), gives the Hamiltonian $ H_s^s $ in schrodinger
picture
\begin{eqnarray}\label{f44.5}
&& H_s^s=H_{1s}^s+H_{2s}^s\nonumber\\
&&H_{1s}^s=2L\sum_{n}\omega_{n}^2B_{n}^\dag
B_{n}\nonumber\\
&&H_{2s}^s=-6\sum_{n,s,t}M_{n,s,-t}B_{n}^\dag B_{s}^\dag
B_{t}B_{n+s-t}-4\sum_{n,s,t}M_{n,s,-t}B_{t}^\dag
B_{s} B_{n}B_{t-s-n}\nonumber\\
&&-\sum_{n,s,t}M_{n,s,t}B_{n}^\dag B_{s}^\dag B_{t}^\dag
B_{-n-s-t}^\dag -4\sum_{n,s,t}M_{n,s,t}B_{n}^\dag
B_{s}^\dag B_{t}^\dag B_{n+s+t}\nonumber\\
&&-\sum_{n,s,t}M_{n,s,t}B_{n}B_{s}B_{t}B_{-n-s-t}
\end{eqnarray}
 The eigenvalues of $ H_{s}^s $ up to the first order
perturbation, can be obtained from the following relation
\begin{eqnarray}\label{f44.6}
&&E_{s_1,...,s_j}=\sum_{i=1}^j\omega_{s_i}+E_{
s_1,...,s_j}^{(1)},\nonumber\\
&&E_{s_1,...,s_j}^{(1)}=\langle
s_1,...,s_j|H_{2s}^s|s_1,...,s_j\rangle=\sum_{i=1}^j\sum_{r\neq
i}\frac{2\gamma\pi^4}{3\omega_{s_i}^2\omega_{s_r}^2 L} s_i^2s_r^2.
\end{eqnarray}
Let $ |\psi(t)\rangle_s $ be the state vector in shrodinger
picture and difine $ \widetilde{|\psi(t)\rangle}=
e^{-iH_{2s}t}|\psi(t)\rangle_s $,then $
\widetilde{|\psi(t)\rangle} $ satisfies the time evolution quation
\begin{equation}\label{f44.61}
i\frac{\partial}{\partial
t}\widetilde{|\psi(t)\rangle}=\widetilde{H}_{2s}(t)\widetilde{|\psi(t)}\rangle,
\end{equation}
and up to the first order perturbation $
\widetilde{|\psi(t)\rangle} $ we have
\begin{equation}\label{f44.62}
\widetilde{|\psi(t)\rangle}=( 1-i\int_0^t d t_1
\widetilde{H}_{2s}(t_1))\widetilde{|\psi(0)\rangle}.
\end{equation}
For example if $ \widetilde{|\psi(0)\rangle} $ is the vacuum
state of string $ |0\rangle $ then
\begin{eqnarray}\label{f44.63}
&&\widetilde{|\psi(t)\rangle}=|0\rangle+i\sum_{
n,l,s}\frac{M_{n,l,s}}{\sqrt{(2L)^4\omega_{n} \omega_{
l}\omega_{s}\omega_{n+l+s}}
}|n,l,s,-n-l-s\rangle \nonumber\\
&&\times \frac{\sin
\frac{(\omega_{n}+\omega_{l}+\omega_{s}+\omega_{n+l+s})}{2}t}{\frac{
\omega_{n}+\omega_{l}+\omega_{s}+\omega_{
n+l+s}}{2}}e^{i\frac{(\omega_{n}+\omega_{
l}+\omega_{s}+\omega_{n+l+s})t}{2}},
\end{eqnarray}
and If $ \widetilde{|\psi(0)\rangle}=|j\rangle $, i.e, the string
is in it's $j$-th mode, for some $ j$ then
\begin{eqnarray}\label{f44.64}
&&\widetilde{|\psi(t)\rangle}=|j\rangle+4i\sum_{
n,s}\frac{M_{n,s,-j}}{\sqrt{(2L)^4\omega_{n}
\omega_{s}\omega_{j}\omega_{j-n-s}}
}|n,s,j-n-s\rangle \nonumber\\
&&\times\frac{sin(\frac{(\omega_{n}+\omega_{s}+\omega_{
j-n-s}-\omega_{j})}{2})t}{\frac{( \omega_{n}+\omega_{s}+\omega_{
j-n-s}-\omega_{j})}{2}}e^{i\frac{( \omega_{n}+\omega_{s}+\omega_{
j-n-s}-\omega_{j})t}{2}}\nonumber\\
&&+i\sum_{ n,l,s}\frac{M_{n,l,s}}{\sqrt{(2L)^4\omega_{n} \omega_{
l}\omega_{s}\omega_{n+l+s}}
}|j,n,l,s,-n-l-s\rangle\nonumber\\
&&\times\frac{sin(\frac{( \omega_{n}+\omega_{l}+\omega_{
s}+\omega_{n+l+s})}{2})t}{\frac{( \omega_{n}+\omega_{
l}+\omega_{s}+\omega_{n+l+s})}{2}}e^{i\frac{(
\omega_{n}+\omega_{l}+\omega_{s}+\omega_{ n+l+s})t}{2}},
\end{eqnarray}
which gives the probability of transition $
|j\rangle\rightarrow|p,q,r\rangle $ after passing a long time
\begin{eqnarray}\label{f44.65}
&&|\langle
p,q,r|\widetilde{\psi(t})\rangle|^2=\frac{32\pi^9\gamma^2 t
}{L^6\omega_{p}\omega _{q}\omega_{r}\omega_{ j}}(jpqr)^2
\delta_{j,p+q+r}
\delta(\omega_{p}+\omega_{q}+\omega_{r}-\omega_{j})\nonumber\\
&&
\end{eqnarray}
 We can also determine $ C_{n}(t) $ up to the first
order aproximation to be
\begin{eqnarray}\label{f44.7}
&&C_{n}(t)=E_{n}e^{i\omega_{n}t}+F_{n}e^{-i\omega_{n}t}\nonumber\\
&&+\sum_{s,j}\frac{L_{-s,-j}B_{s}^\dag B_{j}^\dag
B_{s+j+n}}{\omega_{n}^2-(
\omega_{s}+\omega_{j}-\omega_{s+j-n})^2}e^{i(\omega_{s}+\omega_{j}-\omega_{s+j-n})t}\nonumber\\
&&+\sum_{s,j}\frac{L_{-s,-j}B_{s}^\dag B_{j}^\dag
B_{-s-j-n}}{\omega_{n}^2-(
\omega_{s}+\omega_{j}+\omega_{s+j-n})^2}e^{i(\omega_{s}+\omega_{j}+\omega_{s+j-n})t}\nonumber\\
&&+\sum_{s,j}\frac{L_{-s,j}B_{s}^\dag B_{j}
B_{s+n-j}}{\omega_{n}^2-(
\omega_{s}-\omega_{j}-\omega_{s+j-n})^2}e^{i(\omega_{s}-\omega_{j}-\omega_{s+j-n})t}\nonumber\\
&&+\sum_{s,j}\frac{L_{-s,j}B_{s}^\dag B_{j-s-n}^\dag B_{j}}
{\omega_{n}^2-(
\omega_{s}-\omega_{j}+\omega_{s+j-n})^2}e^{i(\omega_{s}-\omega_{j}+\omega_{s+j-n})t}\nonumber\\
&&+\sum_{s,j}\frac{L_{s,-j}B_{j}^\dag B_{s}B_{n+j-s}}
{\omega_{n}^2-(
\omega_{j}-\omega_{s}-\omega_{s+j-n})^2}e^{i(\omega_{j}-\omega_{s}-\omega_{s+j-n})t}\nonumber\\
&&+\sum_{s,j}\frac{L_{s,-j}B_{j}^\dag B_{s-n-j}^\dag B_{s}}
{\omega_{n}^2-(
\omega_{j}-\omega_{s}+\omega_{s+j-n})^2}e^{i(\omega_{j}-\omega_{s}+\omega_{s+j-n})t}\nonumber\\
&&+\sum_{s,j}\frac{L_{s,j}B_{j}B_{s} B_{n-s-j}}{\omega_{n}^2-(
\omega_{j}+\omega_{s}+\omega_{s+j-n})^2}e^{-i(\omega_{j}+\omega_{s}+\omega_{s+j-n})t}\nonumber\\
&&+\sum_{s,j}\frac{L_{s,j}B_{s+j-n}^\dag B_{j}B_{s} }
{\omega_{n}^2-(
\omega_{j}+\omega_{s}-\omega_{s+j-n})^2}e^{-i(\omega_{j}+\omega_{s}-\omega_{s+j-n})t}
\end{eqnarray}
and the operators $ E_{n} , F_{n} $, can be determined from the
initial conditions
\begin{eqnarray}\label{f44.8}
&&C_{n}(0)=B_{-n}^\dag+B_n,\nonumber\\
&&\dot{C}_{n}(0)=i\omega_{n}B_{-n}^\dag-i\omega_{n}B_{n}.
\end{eqnarray}
\section{Interaction with the quantum vacuum field}
When the quantized vibrating string interacts with the quantum
vacuum of the elctromagnetic field, the total Hamiltonian can be
written as
\begin{equation}\label{f45}
H=H_s+H_F+H'=H_{1s}+H_{2s}+H_F+H',
\end{equation}
where $ H_s=H_{1s}+H_{2s} $, is the string Hamiltonian defined
in(\ref{f43}), $ H_F $ is the Hamiltonian of the vacuum field
 and $ H'$ is the interaction part, defiened by
\begin{eqnarray}\label{E46}
&&H_F=\int d^3x[-\frac{1}{2}\pi_\mu \pi^\mu
+\frac{1}{2}\sum_{k=1}^3 (\partial_k A_\nu) (\partial^k
A^\nu)]\nonumber\\
&&H'=\int d^3xj_\mu A^\mu,
\end{eqnarray}
where $ \pi^\mu=-\frac{\partial A^\mu}{\partial t}$ is canonical
momentum density corresponding to $ A^\mu $ and the Lagrangian
density of the electromagnetic field is[5]
\begin{equation}\label{E47}
\pounds'=-\frac{1}{2}(\partial_\mu A_\nu)(\partial^\mu
A^\nu)-j_\mu A^\mu.
\end{equation}
 Suppose a vibrating medium has the electrical charge
 density $\rho(x)$, and let $ \vec{\eta}(\vec{x},t)$ be the mechanical wave
 propagating in the medium, i.e, $ \vec{\eta}(\vec{x},t)$ is the
 departure from the stable state of an infinitesimal element with center $\vec{x}$, this vibrating
 medium interacts with the quantized electromagnetic field as follows,
 \begin{eqnarray}\label{f32}
 & &H'= \int j_\mu(\vec{x},t)A^\mu(\vec{x},t) d^3x=\int d^3x\rho(\vec{x})[A^0( \vec{x},t)+\vec{\nabla} A^0( \vec{x},t)\cdot\vec{\eta}
(\vec{x},t)\nonumber\\
&-& \frac{\partial\vec{\eta}(\vec{x},t)}{\partial
t}\cdot\vec{A}(\vec{x},t)-(\vec{\eta}(\vec{x},t)\cdot\nabla)\vec{A}(\vec{x},t)\cdot\frac{\partial
\vec{\eta}(\vec{x},t)}{\partial t}+\cdots],
\end{eqnarray}
where $\rho(\vec{x})$ is the charge density before the medium
starts to vibrating. We rewrite $H'$ up to the dipole
approximation so
 \begin{eqnarray}\label{f33}
 & &\int j_\mu(\vec{x},t)A^\mu(\vec{x},t)d^3x\sim\int d^3x
 \rho(\vec{x})(\vec{\nabla}A^0(\vec{x}
,t)+\frac{\partial\vec{A}(\vec{x},t)}{\partial t})\cdot\vec{\eta}(\vec{x},t)\nonumber\\
&=&-\int d^3x\vec{P}(\vec{x},t)\cdot\vec{E}(\vec{x},t),
\end{eqnarray}
where $ \vec{P}(\vec{x},t)$ is the electric dipole density of the
vibrating medium. For the string with a charge density $\sigma $
and length $ L $ stretched on the $ x^1 $ axis which its two ends
are at $x=0$ and $x=L$ we can write the charge density field as
follows
\begin{equation}\label{f34}
\rho(\vec{x})=\sigma\delta(x^2)\delta(x^3)(u(x^1)-u(x^1-L)),
\end{equation}
where $ u$ is the step function, assume that the string vibrates
only in the $x^2$ direction such that $ y(x^1,t)\vec{j}$ be the
corresponding mechanical wave function, so by using (\ref{f33}),
the interaction part of the Hamiltonian can be written as
\begin{eqnarray}\label{f35}
H'(t)&=&-\sigma\int_0^L dx^1y(x^1,t)E^2(x^1,0,0,t)
\end{eqnarray}
By defining $ H_0:=H_{1s}+H_F $ and $ H'':=H_{2s}+H' $ such that $
H=H_0+H'' $, we have in the interaction picture
\begin{eqnarray}\label{f46}
&&y_I(x^1,t)= e^{iH_0t}y(x^1,0)e^{-iH_0t}
= \sum_{n=-\infty}^{+\infty}(B_ne^{-i|\omega_n|t+ik_nx^1}+B_n^\dag e^{i|\omega_n|t-ik_nx^1})\nonumber\\
&&(\pi_y)_I(x^1,t)= e^{iH_0t}\pi_y(x^1,0)e^{-iH_0t} =
i\sum_{n=-\infty}^{+\infty}|\omega_n|(B_n^\dag e^
{i|\omega_n|t-ik_nx^1}-B_n e^{-i|\omega_n|t+ik_nx^1})\nonumber\\
&&
\end{eqnarray}
where $ k_n=\frac{2\pi n}{L} , \omega_n=\frac{2\pi n v}{L}$. The
new annihilation and creation operators $ B_{n}$  and $B_{m}^\dag
$ of the string satisfy the commutation relation (\ref{f44.3}).
For electromagnetic field we can write the following expansions in
the interaction picture
\begin{eqnarray}\label{f30}
A_I^\mu(\vec{x},t)&=&\int\frac{d^3k}{\sqrt{2(2\pi)^3\omega_k}}
\sum_{\lambda=0}^3( b_{k\lambda}^\dag
e^{i\omega_kt-i\vec{k}\cdot\vec{x}}+b_{k\lambda}e^{-i\omega_kt+i\vec{k}\cdot\vec{x}})
\varepsilon^\mu(\vec{k},\lambda),\nonumber\\
\vspace{1.0 cm}
\pi_I^\mu(\vec{x},t)&=&-\int\frac{d^3ki\omega_k}{\sqrt{2(2\pi)^3\omega_k}}
\sum_{\lambda=0}^3(b_{k\lambda}^\dag
e^{i\omega_kt-i\vec{k}\cdot\vec{x}}-b_{k\lambda}e^{-i\omega_kt+i\vec{k}\cdot\vec{x}})
\varepsilon^\mu(\vec{k},\lambda),\nonumber\\
\vec{E_I}(\vec{x},t)&=&i\sum_{\lambda=1}^2\int
d^3k\sqrt{\frac{\omega_k}{2(2\pi)^3}}
\vec{\varepsilon}(\vec{k},\lambda)(b_{k\lambda}e^{-i\omega_kt+i\vec{k}\cdot\vec{x}}-
b_{k\lambda}^\dag e^{i\omega_kt-i\vec{k}\cdot\vec{x}}),\nonumber\\
&&
\end{eqnarray}
and the creation and anihilation operators $ b_{k\lambda}$ and $
b_{k\lambda}^\dag $ satisfy the comutation relation
\begin{equation}\label{f31}
[b_{k\lambda},b_{k'\lambda'}^\dag]=\delta_{\lambda\lambda'}\delta(
\vec{k}-\vec{k'}),
\end{equation}
one can easily obtain the Hamiltonian $ (H_0)_I $ in the
interaction picture
\begin{equation}\label{f48}
( H_0)_I=H_0=L\sum_{n=-\infty}^{+\infty}\omega_n^2B_n^\dag
B_n+\int d^3k\sum_{\lambda=1}^2 \omega_k b_{k\lambda}^\dag
b_{k\lambda}.
\end{equation}
If we ignore $ H_{2s} $ in the Hamiltonian $ H'' $, i.e, $ H''=H'
$, and write $ H' $ in the interaction picture as
\begin{equation}\label{f49}
H'_I(t)= -\sigma \int_{0}^{L}y_I( x^1,t)E_I^2(x^1,0,0,t)dx^1,
\end{equation}
then by substituting $ y_I(x^1,t) $ from (\ref{f46}) and $ E_I^2(
x^1,0,0,t)$ from (\ref{f30}) one can obtain $ H'_I(t) $ and
calculate the probability amplitude for various transitions. Up
to the first order perturbation, the evolution operator
$U^{(1)}$, in interaction picture, is
\begin{equation}\label{f50}
U^{(1)}(+\infty,-\infty)=1-i\int_{-\infty}^{+\infty}H'_I(t)dt,
\end{equation}
for example the probability amplitude for transition from
 $ |0\rangle_F \otimes|m\rangle_s $ to\\
  $|\vec{q},r\rangle_F \otimes|0\rangle_s $ \\ is
\begin{equation}\label{f51}
\frac{-i\sigma  \sqrt{\omega_q}}{2\sqrt{2\pi L|\omega_m|}}
\varepsilon^2(\vec{q},r)\delta(\omega_q-|\omega_m|)\frac{e^{-iq^1L}-1}{k_m-q^1},
\end{equation}
where  $\vec{q} $ and $r$ are momentum and polarization of the
created photon respectively, $ m $ is the quantum number for
string quanta, $ |0\rangle_F $ and $ |0\rangle_s $ are
electromagnetic and string vacuums respectively. In the second
order perturbation we have
\begin{equation}\label{f53}
U^{(2)}(+\infty,-\infty)=1-i\int_{-\infty}^{+\infty}H'_I(t)d t
-\frac{1}{2}\int_{-\infty}^{+\infty}dt_1\int_{-\infty}^{+\infty}dt_2
T(H'(t_1)H'(t_2),
\end{equation}
by using of wick theorem [5] the last term can be written as
\begin{eqnarray}\label{f54}
&-&\frac{1}{2}\int_{-\infty}^{+\infty}dt_1\int_{-\infty}^{+\infty}dt_2
T(H'(t_1)H'(t_2)=\nonumber\\
&&-\frac{\sigma^2}{2}\int_{-\infty}^{+\infty} dt_{1}
\int_{0}^{L}dx^{1}
\int_{-\infty}^{+\infty} dt_{2} \int_{0}^{L} dz^{1} \nonumber\\
&\times&:\ y_I(x^1,t_1) E_I^{2}(x^1,0,0,t_1) y_I(z^1,t_2)E_I^{2}(z^1,0,0,t_2):\nonumber\\
&+&:E_{I}^{2}(x^{1},0,0,t_{1})E_{I}^{2}(z^{1},0,0,t_{2}):\langle 0|y_{I}(x^{1},t_{1})y_{I}(z^{1},t_{2})|0\rangle \nonumber\\
&+ &: y_{I}(x^1,t_1)y_{I}(z^1,t_2) :\langle 0 |E_{I}^{2}(x^1,0,0,t_1)E_{I}^{2}(z^1,0,0,t_2)|0\rangle \nonumber\\
&+& \langle 0 |y_{I}(x^1,t_1) y_{I}(z^1,t_2)|0\rangle \langle 0
|E_{I}^{2}(x^1,0,0,t_1)E_{I}^{2}(z^1,0,0,t_2)|0\rangle\}.
\end{eqnarray}
where $ :\hspace{0.5cm}  : $ denote the normal ordering. The
second term under integral has no efect on the probability
amplitude of those transitions that initial and final string
states are different,The third term under integral has no efect on
the probability amplitude of those transitions that initial and
final photon states are different, also the last term has no
effect on the probability amplitude of those transitions that
initial and final states both string and electromagnetic field
are different. By substituting $ y_I(x^1,t)$ from (\ref{f46}) and
$E_I^3(x^1,0,0,t) $ from (\ref{f30}) into the first term in above
equation, one can calculate the probability amplitude for some
special transitions. For example
probability amplitude for the transition\\
 $ |m\rangle\otimes
|\vec{p},r_1\rangle_F\rightarrow|n\rangle\otimes|\vec{q},r_2\rangle_F
$ is
\begin{eqnarray}\label{f55}
&&-\frac{\sigma^2 \sqrt{\omega_q\omega_p}}{16\pi L
\sqrt{|\omega_m||\omega_n|}}
\varepsilon^2(\vec{p},r_1)\varepsilon^2(\vec{q},r_2)
\delta(\omega_q-|\omega_m|)\delta(\omega_p-|\omega_n|)
\frac{(e^{-iq^1L}-1)(e^{ip^1L}-1)}{(k_n-p^1)(k_m-q^1)}\nonumber\\
&&,
\end{eqnarray}
 If we keep $ H_{2s}$ in $ H''$ ,then $ H''=H'+H_{2s} $ and up to
the first order  perturbation, we have
\begin{equation}\label{f57}
U^{(1)}(+\infty,-\infty)=1-i \int_{-\infty}^{+\infty}dt
\int_{0}^{L}dx^1[\frac{\gamma}{24}(\frac{\partial
y_I(x^1,t)}{\partial x^1})^4+\sigma y_I(x^1,t)E_I^2(x^1,0,0,t)],
\end{equation}
in this case, the term $ \frac{1}{24}(\frac{\partial
y(x^1,t)}{\partial x^1})^4
 $ has no effect on the probability amplitude of those transitions that
the initial and the final photon states are different. \\
 In the second order perturbation, we can write
 \begin{eqnarray}\label{f58}
&-&\frac{1}{2}\int_{-\infty}^{+\infty}dt_1\int_{-\infty}^{+\infty}dt_2
T(H'(t_1)H'(t_2) \nonumber\\
&=&-\frac{1}{2}\int_{-\infty}^{+\infty}d
 t_1\int_{0}^L dx^1\int_{-\infty}^{+\infty}d t_2\int_0^L dz^1 \nonumber\\
 &\times&\{\frac{\gamma^2}{576}T [:(\frac{\partial y(x^1,t_1)}{\partial
 x^1})^4 : :( \frac{\partial y(z^1,t_2)}{\partial z^1})^4:]\nonumber\\
 &+&\frac{\gamma\sigma}{24}T [: (\frac{\partial y(x^1,t_1)}{\partial
 x^1})^4 : y(z^1,t_2) E^2(z^1,0,0,t_2):]\nonumber\\
 &+&\frac{\gamma\sigma}{24}T[:y(x^1,t_1) E^2(x^1,0,0,t_1) : : ( \frac{\partial y (z^1,t_2)}{\partial
 z^1})^4 ]\nonumber\\
 &-&\sigma^2 T [ :y(x^1,t_1)E^2(x^1,0,0,t_1): : y
 (z^1,t_2)E^2(z^1,0,0,t_2):]\}.
 \end{eqnarray}
 The first term under integral has no effect on those probability amplitude that
 the initial photon state is different from final photon state.the
 effect of fourth term under integral is similar to (\ref{f54})and has no effect on
 some of transitions. For example transition
$|m\rangle_s\otimes|l\rangle_s|f\rangle_s\otimes|0\rangle_F\longrightarrow|n\rangle_s\otimes|j\rangle_s\otimes|\vec{p},r\rangle_F$\\
up to the second order perturbation, has the following
probability amplitude
\begin{eqnarray}\label{f59}
&& \frac{-\gamma\sigma
\varepsilon^3(\vec{p},r)}{4\sqrt{\omega_{l}\omega_{f}\omega_{m}\omega_{n}\omega_{j}}}\sqrt{\frac{(2\pi)^9\omega_p}{L^{11}}}\nonumber\\
 &&\times\{mlnj
\delta(\omega_{n}+\omega_{j}-\omega_{m}-\omega_{l})\delta(\omega_{f}-\omega_p)
\delta_{m+l,n+j}\frac{e^{-ip^1L}-1}{p^1-\frac{2\pi f}{L}}\nonumber\\
&&+mfnj\delta(\omega_{n}+\omega_{j}-\omega_{m}-\omega_{f})\delta(\omega_{l}-\omega_p)
\delta_{m+f,n+j}\frac{e^{-ip^1L}-1}{
p^1-\frac{2\pi l}{L}}\nonumber\\
&&+lfnj\delta(\omega_{n}+\omega_{j}-\omega_{l}-\omega_{f})\delta(\omega_{m}-\omega_p)
\delta_{l+f,n+j}\frac{ e^{-ip^1L}-1}{p^1-\frac{2\pi m}{L})}\},
\end{eqnarray}
which is only the effect of the second and third term
in(\ref{f58}).That is  only the nonlinear term $
\frac{\gamma}{24}(\frac{\partial y}{\partial x})^4 $ in
Hamiltonian (\ref{f39})has a nonzero effect in this probability.
Also the probability amplitude for transition
$|m\rangle_s\otimes|l\rangle_s\otimes|f\rangle_s\otimes
|j\rangle_s\otimes|0\rangle_F\longrightarrow|n\rangle_s\otimes|\vec{p},r\rangle_F$
is
\begin{eqnarray}
&& \frac{-\gamma\sigma
\varepsilon^3(\vec{p},r)}{4\sqrt{\omega_{l}\omega_{f}\omega_{m}\omega_{n}\omega_{j}}}\sqrt{\frac{(2\pi)^9\omega_p}{L^{11}}}\nonumber\\
&&\times\{mlnf\delta(\omega_{n}-\omega_{f}-\omega_{m}-\omega_{l})\delta(\omega_{j}-\omega_p)
\delta_{m+l+f,n}\frac{e^{-ip^1L}-1)}{ p^1-\frac{2\pi j}{L}}\nonumber\\
&&+(mlnj\delta(\omega_{n}-\omega_{l}-\omega_{m}-\omega_{j})\delta(\omega_{f}-\omega_p)
\delta_{m+l+j,f}\frac{e^{-ip^1L}-1}{ p^1-\frac{2\pi f}{L}}\nonumber\\
&&+lfnj\delta(\omega_{n}-\omega_{j}-\omega_{l}-\omega_{f})\delta(\omega_{m}-\omega_p)
\delta_{l+f+j,n}\frac{e^{-ip^1L}-1}{p^1-\frac{2 \pi m}{L}}\nonumber\\
&&+mfnj\delta(\omega_{n}-\omega_{j}-\omega_{m}-\omega_{f})\delta(\omega_{l}-\omega_p)\delta_{m+f+j,n}\frac{e^{-ip^1L}-1}{p^1-\frac{2
\pi l}{L}}\}
\end{eqnarray}
which is only due to the efect of nonlinear term $
\frac{\gamma}{24}(\frac{\partial y}{\partial x^1})^4$
\section{The effect of vacuum field and radiation reaction}
In this section we use the coulomb gauge for finding the radiation
reaction effect, for this purpose let us write the free part of
the electromagnetic field as
\begin{equation}\label{f60}
H_F=\frac{1}{2}\int d^3x(
(\vec{E}^\|)^2+(\vec{E}^\bot)^2+2\vec{E}^\bot.\vec{E}^\|+\vec{B}^2)
\end{equation}
where $ \vec{E}^\bot= -\frac{\partial \vec{A}}{\partial t}$ and $
\vec{E}^\|=-\vec{\nabla} A^0 $, $H_{F}$ can be rewritten simply as
\begin{equation}\label{f61}
H_F= \frac{1}{2}\int d^3x (
(\vec{E}^\bot)^2+\vec{B}^2)+\frac{1}{2}\int d^3x j_0(
\vec{x},t)A^0( \vec{x},t),
\end{equation}
where the last term is the coulomb interaction of the string with
itself which clearly is divergent and  we may ignore it because
we are not interested in self interaction effects. So the
effective Hamiltonian can be considered to be
\begin{equation}\label{f62}
H=H_F=\frac{1}{2}\int
d^3x((\vec{E}^\bot)^2+\vec{B}^2)=\sum_{\lambda=1}^2 \omega_k(
a_{k\lambda}^\dag(t)a_{k\lambda}(t)+\frac{1}{2}),
\end{equation}
 and accordingly the total Hamiltonian is
 \begin{eqnarray}\label{f63}
 & &H=H_s+H_F+H'=\nonumber\\
 &=&H_s+\sum_{\lambda=1}^2\omega_k( a_{k\lambda}^\dag(t)
 a_{k\lambda}(t)+\frac{1}{2})-\int d x^3 \vec{j}(
 \vec{x},t).\vec{A}( \vec{x},t),
\end{eqnarray}
 where $ H_s $ is defined in (\ref{f39}). The interaction part of the Hamiltonian (\ref{f32})
   up to the electric dipole approximation is given
 by(\ref{f35})which gives the Heisenberg equation for $ y(x^1,t) $ as
\begin{equation}\label{f64}
\frac{\partial^2 y}{\partial t^2}-v^2\frac{\partial^2 y}{\partial
(x^1)^2}=\frac{\gamma}{2}( \frac{\partial y}{\partial
x^1})^2\frac{\partial^2 y}{\partial (x^1)^2}+\sigma
E^2(x^1,0,0,t).
\end{equation}
The vector potential $ \vec{A} $ and
 transverse electrical field are defined by [1]
\begin{eqnarray}\label{f65}
 \vec{A}( \vec{x},t)&=&\int \frac{d^3k}{\sqrt{2(2\pi)^3\omega_k}}\sum_{\lambda=1}^2(
 a_{k\lambda}(t)e^{+i\vec{k}.\vec{x}}+a_{k\lambda}^\dag(t)e^{-i\vec{k}.\vec{x}})\vec{\varepsilon}(
 \vec{k},\lambda)\nonumber\\
 \vec{E}^\bot( \vec{x},t)&=&\int \frac{d^3ki\omega_k}{\sqrt{2(2\pi)^3\omega_k}}\sum_{\lambda=1}^2(
 a_{k\lambda}(t)e^{+i\vec{k}.\vec{x}}-a_{k\lambda}^\dag(t)e^{-i\vec{k}.\vec{x}})
 \vec{\varepsilon}(\vec{k},\lambda),
 \end{eqnarray}
where the time dependence of $ a_{k\lambda}(t)$ is not simply as $
a_{k\lambda}e^{-i\omega_k t}$ and must be specified by Heisenberg
equation so
\begin{eqnarray}\label{f66}
\dot{a}_{k\lambda}(t)&=&i[ H,a_{k\lambda}(t)]=i[H_s+H_F+H']\nonumber\\
&=&-i\omega_k
a_{k\lambda}(t)-\sigma\sqrt{\frac{\omega_k}{2(2\pi)^3}}\varepsilon^2(
\vec{k},\lambda)\int_0^L d z y(z,t) e^{-ik^1z},
\end{eqnarray}
where we have used the commutation relations
$[a_{k'\lambda'}(t),a_{k\lambda}^\dag(t)]=\delta_{\lambda\lambda'}\delta(\vec{k}-\vec{k'})$.
 A formal solution for above equation can be written as
\begin{equation}\label{f67}
a_{k\lambda}(t)=a_{k\lambda}(0)e^{-i\omega_k
t}-\sigma\sqrt{\frac{\omega_k}{2(2\pi)^3}}\varepsilon^2(
\vec{k},\lambda)\int_0^t d t'e^{i\omega_k(t-t')}\int_0^L d z
y(z,t')e^{-ik^1z},
\end{equation}
now by substituting  $ a_{k\lambda}(t)$ from  equation (\ref{f67})
in electrical field expression (\ref{f65}), one obtains the
electrical field as the sum of two parts, the first part is
nothing but the vacuum field which is
\begin{equation}\label{f68}
E_0^2(x^1,0,0,t)= i\int
d^3k\sqrt{\frac{\omega_k}{2(2\pi)^3}}\sum_{\lambda=1}^2(
a_{k\lambda}(0)e^{-i\omega_k t+ik^1
x^1}-a_{k\lambda}^\dag(0)e^{i\omega_k t-i k^1 x^1})\varepsilon^2(
\vec{k}\lambda),
\end{equation}
the second part is the radiation reaction field
\begin{eqnarray}\label{f69}
&&E_{RR}^2(x^1,0,0,t)=-\frac{\sigma\pi}{(2\pi)^3}\int_0^{2\varphi}d
\varphi \int_0^\pi d \theta \sin\theta
\sum_{\lambda=1}^2(\varepsilon^2(
\vec{k},\lambda))^2\nonumber\\
&\times & \int_0^t d t'\int_0^L d z
y(z,t')\frac{\partial^3}{\partial
(t')^3}\delta(t-t'+(x^1-z)\cos\theta).
\end{eqnarray}
 Integration by parts respect to $ t'$ and then inserting $  \sum_{\lambda=1}^2 ( \varepsilon^2(
\vec{k},\lambda))^2=1-\sin^2\theta \cos^2\varphi $ and doing
integrals with respect to $ \theta , \varphi $, we at last come
to the following relation for radiation reaction component
\begin{equation}\label{f70}
E_{RR}^2(x^1,0,0,t)=\frac{\sigma}{\pi}\int_0^L d z
\sum_{m=0}^\infty\frac{(m+1)(x^1-z)^{2m}}{(2m)!(2m+1)(
2m+3)}\frac{\partial^{2m+3}}{\partial t^{2m+3}}y(z,t),
\end{equation}
the first term, i.e, $ m=0 $ gives the first order approximation
of $ E_{RR}^2 $
\begin{equation}\label{f71}
E_{RR}^2(x^1,0,0,t)=\frac{\sigma}{3\pi}\int_0^L d
z\frac{\partial^3 y(z,t)}{\partial t^3},
\end{equation}
that may be compared with radiation reaction field due to one
dimensional harmonic oscillator [1]. Using (\ref{f71}), the
Heisenberg equation (\ref{f64}) can be written like this
\begin{equation}\label{f72}
\frac{\partial^2 y}{\partial t^2}-v^2\frac{\partial^2 y}{\partial
(x^1)^2}=\frac{\gamma}{24}\frac{\partial^2 y}{\partial (x^1)^2}(
\frac{\partial y}{\partial x^1})^2+\frac{\sigma^2}{3\pi}\int_0^L d
z \frac{\partial^3 y(z,t)}{\partial t^3}+\sigma E_0^2(x^1,0,0,t).
\end{equation}

\end{document}